# Systematic Study of Directed Flow at RHIC Energies

Alice C. Mignerey, University of Maryland, for the PHOBOS Collaboration

B.Alver[4], B.B.Back[1], M.D.Baker[2], M.Ballintijn[4], D.S.Barton[2], R.R.Betts[6], A.A.Bickley[7], R.Bindel[7], A.Budzanowski[3], W.Busza[4], A.Carroll[2], Z.Chai[2], V.Chetluru[6], M.P.Decowski[4], E.Garcia[6], T.Gburek[3], N.George[2], K.Gulbrandsen[4], S.Gushue[2], C.Halliwell[6], J.Hamblen[8], G.A.Heintzelman[2], C.Henderson[4], I.Harnarine[6], D.J.Hofman[6], R.S.Hollis[6], R.Holynski[3], B.Holzman[2], A.Iordanova[6], E.Johnson[8], J.L.Kane[4], N.Khan[8], W.Kucewicz[6], P.Kulinich[4], C.M.Kuo[5], W.Li[4], W.T.Lin[5], C.Loizides[4], S.Manly[8], A.C.Mignerey[7], R.Nouicer[2,6], A.Olszewski[3], R.Pak[2], I.C.Park[8], C.Reed[4], L.P.Remsberg[2], M.Reuter[6], E.Richardson[7], C.Roland[4], G.Roland[4], L.Rosenberg[4], J.Sagerer[6], P.Sarin[4], P.Sawicki[3], I.Sedykh[2], W.Skulski[8], C.E.Smith[6], M.A.Stankiewicz[2], P.Steinberg[2], G.S.F.Stephans[4], A.Sukhanov[2], A.Szostak[2], J.-L.Tang[5], M.B.Tonjes[7], A.Trzupek[3], C.Vale[4], G.J.van-Nieuwenhuizen[4], S.S.Vaurynovich[4], R.Verdier[4], G.I.Veres[4], p.Walters[8], E.Wenger[4], D.Willhelm[2], F.L.H.Wolfs[8], B.Wosiek[3], K.Wozniak[3], A.H.Wuosmaa[1], S.Wyngaardt[2], and B.Wyslouch[4]

[1]Argonne National Laboratory, Argonne, IL 60439-4843, USA
[2]Brookhaven National Laboratory, Upton, NY 11973-5000, USA
[3]Institute of Nuclear Physics PAN, Krakow, Poland
[4]Massachusetts Institute of Technology, Cambridge, MA 02139-4307, USA
[5]National Central University, Chung-Li, Taiwan
[6]University of Illinois at Chicago, Chicago, IL 60607-7059, USA
[7]University of Maryland, College Park, MD 20742, USA
[8]University of Rochester, Rochester, NY 14627, USA

Abstract:

Directed flow, $v_1$, of charged hardons has been measured in Au-Au collisions at RHIC for center-of-mass energies $\sqrt{s_{NN}}$ = 19.6, 130, 62.4, and 200 GeV using the PHOBOS detector. The large acceptance of PHOBOS for charged particles allows measurements over the full range of pseudorapidity $|\eta|$ <5.4.  The results for a symmetric subevent method are shown at all four energies.  Comparison is made to a mixed harmonic method for the highest energy, and compared to similar results from the STAR collaboration.

The distribution of charged particles produced with respect to the plane of the reaction was first quantified for heavy ion reactions at relativistic energies at the Lawrence Berkeley National Laboratory Bevalac, and later at the Brookhaven National Laboratory AGS and the CERN SPS.  The so-called flow has been viewed as a sensitive measure of the equation of state of nuclear matter at those energies.  With the advent of heavy ion experiments at RHIC, these studies have been extended to ultrarelativistic energies and serve as a critical test of hydrodynamic models.  Many measurements have been made of the elliptic flow or $v_2$ component of the distribution; only recently have measurements of the deflection from the beam direction in the plane of the reaction been made at RHIC energies.  This directed flow or $v_1$ is much smaller than that observed at the lower energies, but is nonetheless quantifiable.  This paper describes the directed flow measurements for the Au-Au system at four collision energies, $\sqrt{s_{NN}}$ = 19.6, 130, 62.4, and 200 GeV, made using the PHOBOS detector at RHIC.

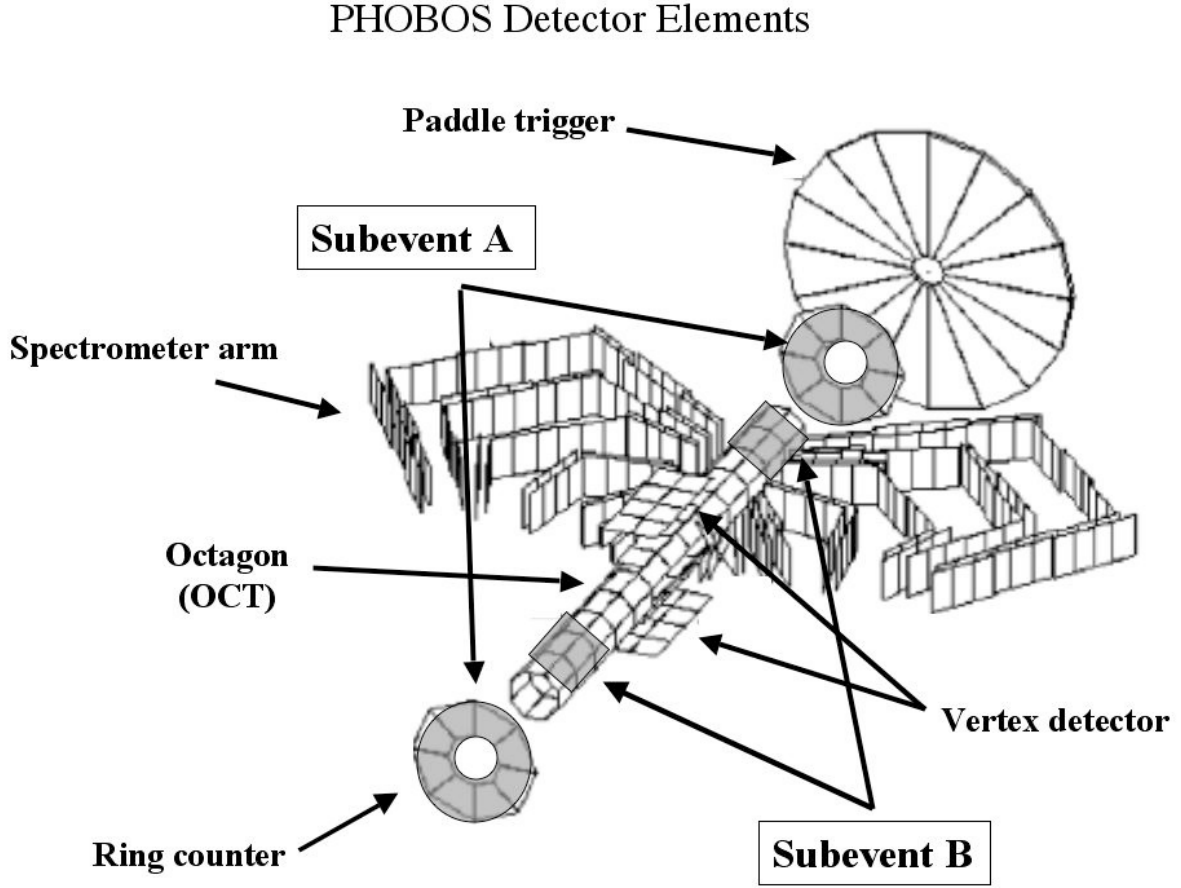

Figure 1: A diagram of the elements of the PHOBOS detector. The regions relevant to the determination of the reaction plane are shaded.

Directed flow is quantified by measuring the first harmonic, $v_1$, of the Fourier decomposition of the azimuthal distribution of particle angles $\phi$, where $\psi_R$ is the true reaction plane angle defined by the impact parameter and beam axis,

$$\frac{dN}{d(\phi - \psi_R)} \sim 1 + \sum_n 2 v_n \cos[n(\phi - \psi_R)]. \qquad [1]$$

In practice, the event plane $\psi_n$ is used, as determined by

$$\psi_n = \left( \tan^{-1} \frac{\sum_i w_i \sin(n\phi_i)}{\sum_i w_i \cos(n\phi_i)} \right) / n, \qquad [2]$$

where

$\psi_n$ = reaction plane angle derived from $n^{th}$ Fourier component, $n = 1$ for $v_1$
$\phi_i$ = angle (in laboratory frame) of the $i^{th}$ hit
$w_i$ = corrective weights applied to $i^{th}$ hit.

A diagram of the PHOBOS detector is shown in Fig. 1. A full description of the detector can be found in [1]. The relevant segments of the detector are highlighted and labeled. This is a hit based analysis using the silicon sensors in the Octagon (OCT) and Ring (RINGS) detectors. The paddle scintillator detectors are used to determine the event centrality. Symmetric subevents are chosen to ensure momentum conservation between the two pseudorapidity ($\eta$) hemispheres. The subevent regions used in the event plane calculations are well separated to minimize non-flow correlations and are located in the OCT ($1.5 < |\eta| < 3$) and RINGS ($3 < |\eta| < 5$) subdetectors. The reaction plane defined by the OCT subevents is used to find $v_1$ in the RINGS region, while that defined by the RINGS subevents is used to measure $v_1$ in the OCT.

The following corrections have been made to the raw hit distributions and the resulting flow signal:

- A weighting matrix is applied to correct for phase space differences between detector pads and to account for detector occupancy.
- The reaction plane resolution correction is found by comparing the symmetric $\eta<0$ with $\eta>0$ regions [2].
- A suppression correction based on a reconstructed known Monte Carlo flow signal is applied. This accounts for a reduction in the measured flow due to background particles that carry no flow information.

The resulting directed flow $v_1$ is shown in Fig. 2 for the four Au-Au energies. Note that the magnitude is much larger for the two lower energies and the $v_1$ axes for these and the two higher energies are different. There is a directed flow observed at all energies, with a turn around or inversion seen at the highest $|\eta|$ in all cases. This is most pronounced for the lowest energies, where the flow changes sign at high $|\eta|$. A direct comparison of the four energies can be made by reflecting about $\eta = 0$ and transforming to reference frame $\eta' = |\eta| - y_{beam}$. This gives the results in a reference frame approximately equal to the rest frame of one of the beams. Figure 3(top) shows the data for each energy from Fig. 2 reflected and averaged. For clarity, only the statistical error bars are included. It is now apparent that the inversion of sign takes place around the beam rapidity and is probably due to spectator protons. There is a remarkable universal scaling in terms of $\eta'$, reminiscent of that which is observed in the elliptic flow, shown in Fig. 3(bottom) (taken from [3].) This extended longitudinal scaling seen in directed flow and elliptic flow was previously observed in the pseudorapidity distribution of charged particles [4].

The data were also analyzed using a mixed harmonic technique [5], which accounts for possible non-flow effects by comparing the directed flow signal obtained with respect to two different event planes, as defined by Eq. [2]. The event plane, $\psi_2$, is determined using the elliptic flow information. The directed flow signal perpendicular to $\psi_2$ is subtracted from that which is in the plane of $\psi_2$. Since a directed flow signal should lie in the reaction plane, any out-of-plane signal is interpreted as due to non-flow effects.

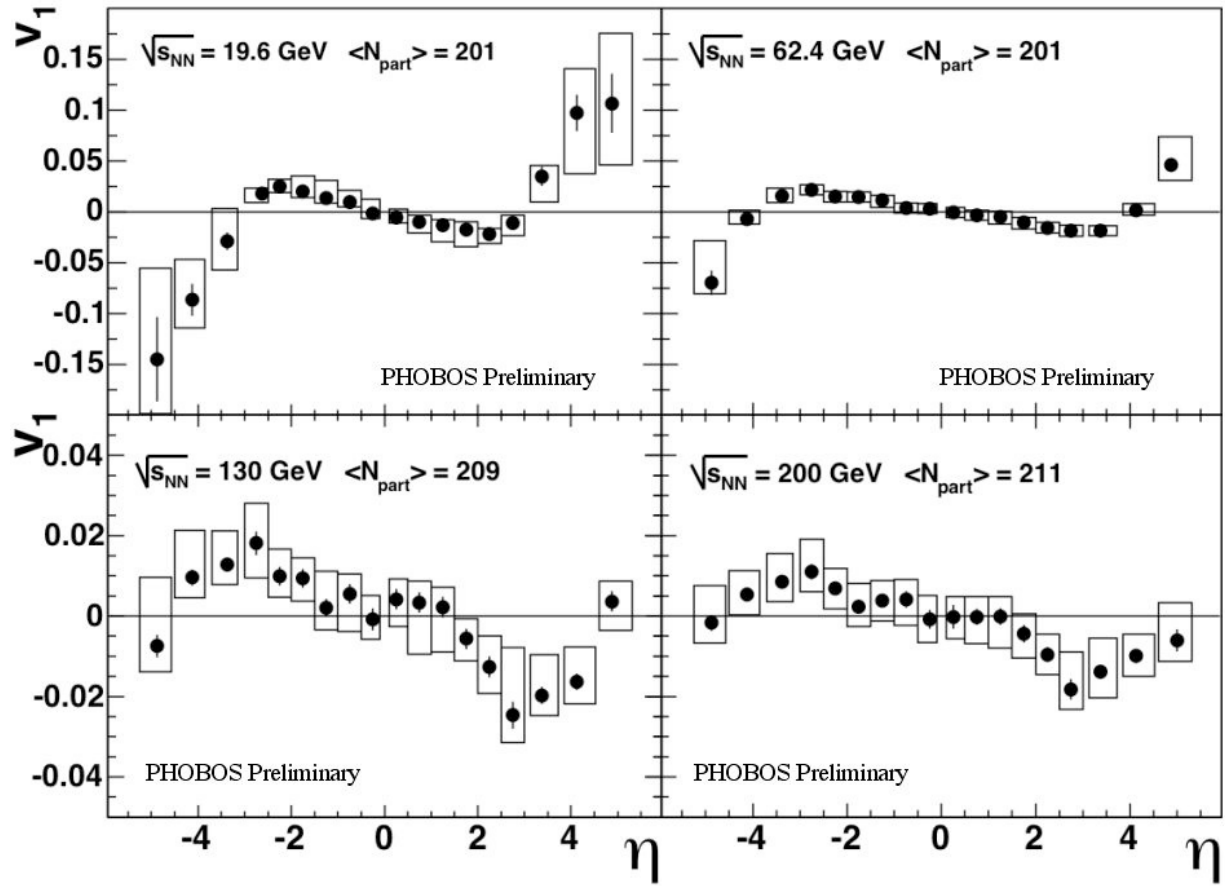

Figure 2. Directed flow in Au-Au collisions at $\sqrt{s_{NN}}$ =19.6, 62.4, 130, and 200 GeV, as a function of pseudorapidity, for the centrality range of 0-40%. The lines represent one-sigma statistical errors and the boxes are 90% C.L. systematic errors.

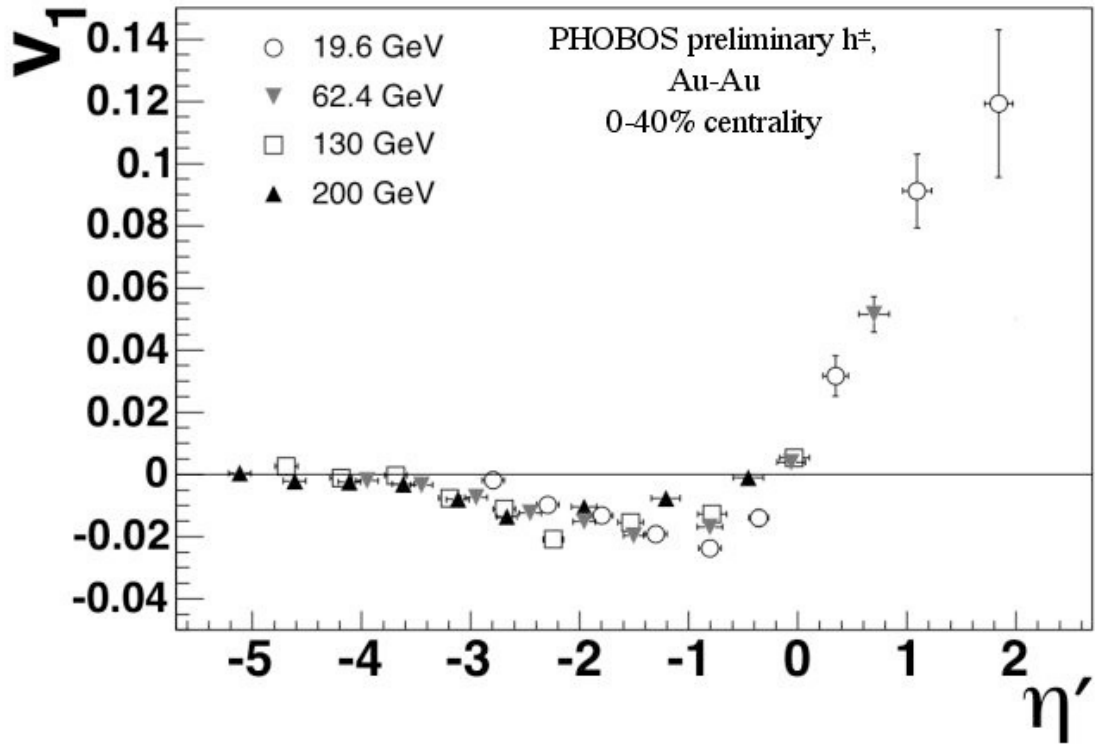
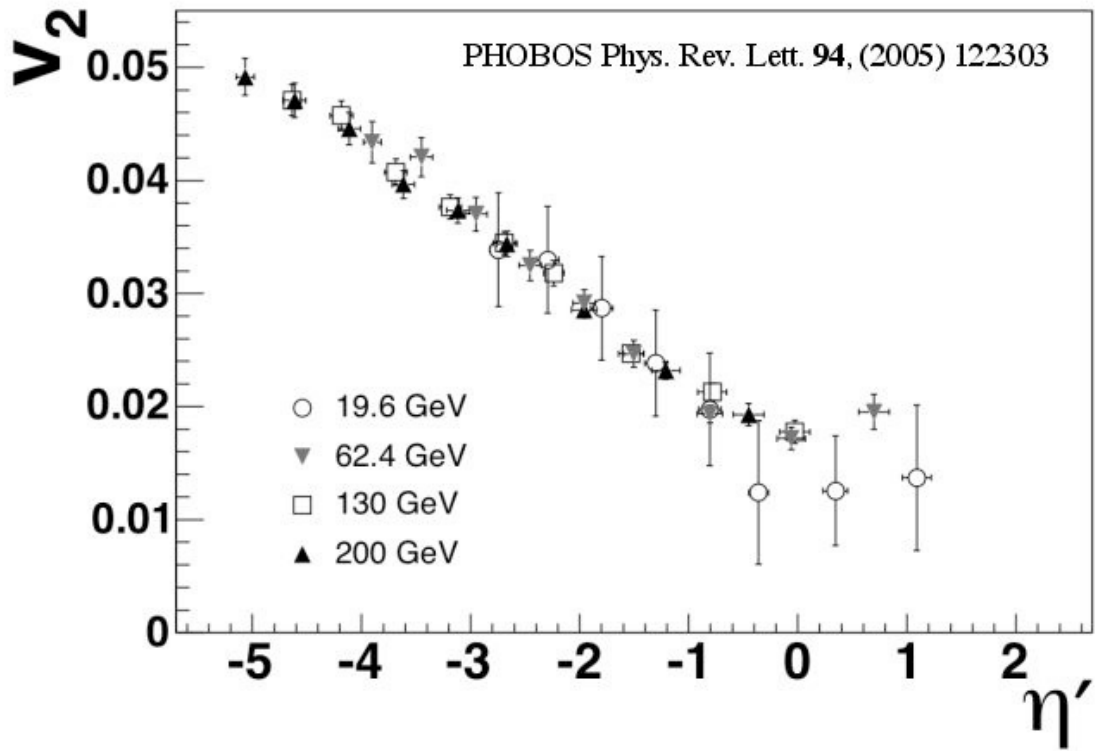

Figure 3. Directed flow (top) and elliptic flow (bottom) as a function of $\eta' = |\eta| - y_{beam}$, averaged over the centrality range of 0-40%, for Au-Au collisions at $\sqrt{s_{NN}}$ =19.6, 62.4, 130, and 200 GeV. Only the one sigma statistical errors are shown.

The mixed harmonic results for all four energies are in agreement with that obtained using the subevent technique. Figure 4 shows a comparison between the two techniques for the 200-GeV data, which agree within the statistical errors. Also shown are results from STAR using the mixed harmonic method [5]. Both PHOBOS and STAR show that the flow is consistent with zero in the range of |η| < 1. There is a slight discrepancy for the larger |η| values; PHOBOS indicates a flattening or turn over, although the centrality regions for the PHOBOS and STAR data are slightly different and the error bars are too large to reach a definitive conclusion. The PHOBOS detector also has a lower energy cut-off in the detected particles at large |η| due to the length of the beryllium beam pipe.

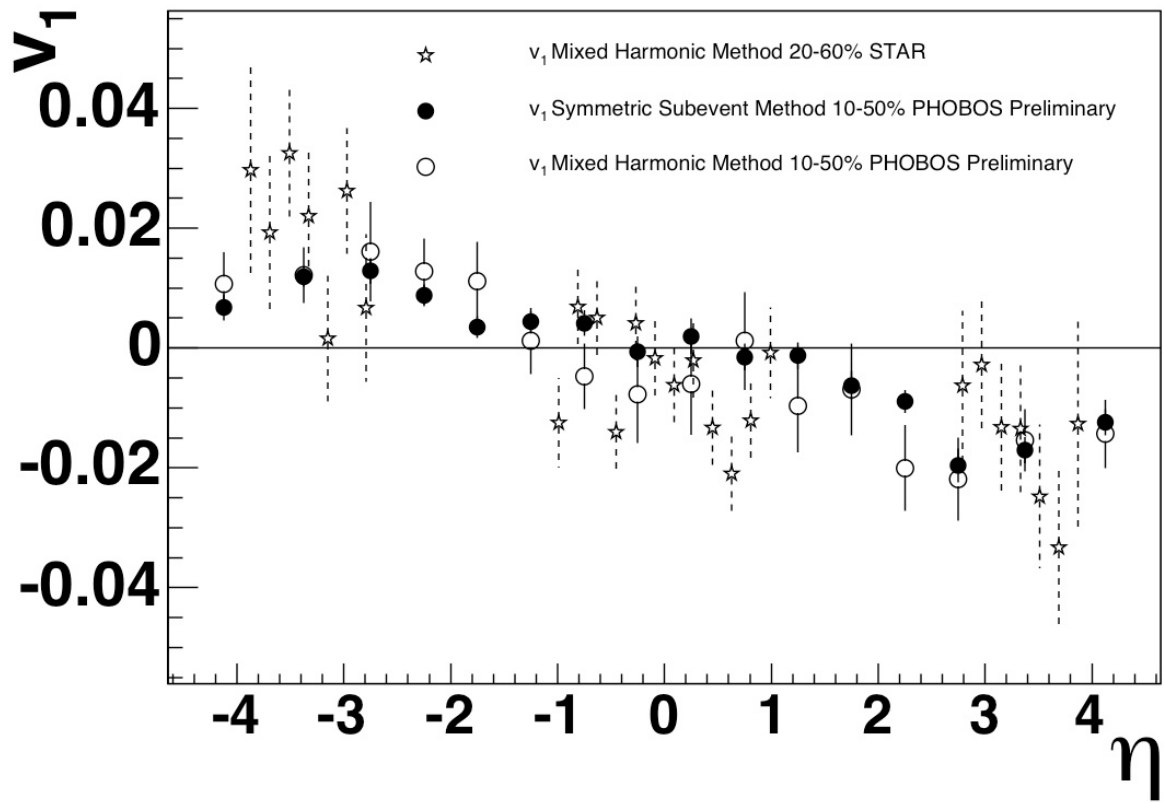

Figure 4: The directed flow $v_1$ as a function of pseudorapidity, for the Au-Au system measured at $\sqrt{s_{NN}}$ =200 GeV. The data from STAR is from [5], and the centrality ranges are as noted.

PHOBOS has measured the directed flow of charged hadrons over the pseudorapidity range |η| <5.4 for $\sqrt{s_{NN}}$ = 19.6, 130, 62.4, and 200 GeV. When transformed into a rest frame approximately equal to that of one of the beams, the data for the four energies follows a universal curve, as seen for the elliptic flow. In this representation, it is evident that the change in the sign of the flow signal as a function of pseudorapidity occurs near the beam rapidity and is probably due to the detection of spectator protons.

Acknowledgments:

This work was partially supported by U.S. DOE grants DE-AC02-98CH10886, DE-FG02-93ER40802, DE-FC02-94ER40818, DE-FG02-94ER40865, DE-FG02-99ER41099, and W-31-109-ENG-38, by U.S. NSF grants 9603486, 0072204, 0245011, by Polish KBN grant 1-P03B-062-27(2004-2007), by NSC of Taiwan Contract NSC 89-2112-M-008-024, and by Hungarian OTKA grant (F 049823).